\documentclass[floatfix,twocolumn,pre]{revtex4}%
\usepackage{amsmath}
\usepackage{amsfonts}
\usepackage{amssymb}
\usepackage{graphicx}
\usepackage{hyperref}%
\begin{document}
\title{Socioeconomic Networks with Long-Range Interactions}
\author{Rui Carvalho}
\email{rui.carvalho@ucl.ac.uk}
\affiliation{Centre for Advanced Spatial Analysis, 1-19 Torrington Place, University
College London, WC1E 6BT United Kingdom}
\author{Giulia Iori}
\email{g.iori@city.ac.uk}
\affiliation{Department of Economics, School of Social Science City University, Northampton
Square, London EC1V 0HB United Kingdom}

\begin{abstract}

We study a modified version of a model previously proposed by Jackson and
Wolinsky to account for communicating information and allocating goods in
socioeconomic networks. In the model, the utility function of each node is
given by a weighted sum of contributions from all accessible nodes. The 
weights, parametrized by the variable $\delta$, decrease with distance. We
introduce a growth mechanism where new nodes attach to the existing network
preferentially by utility. By increasing $\delta$, the network structure
evolves from a  power-law to an exponential degree distribution,
passing through a regime characterised by shorter average path length, lower
degree assortativity and higher central point dominance. In the second part of
the paper we compare different network structures in terms of the average
utility received by each node. We show that power-law networks provide higher
average utility than Poisson random networks. This provides a possible
justification for the ubiquitousness of scale-free networks in the real world.

\end{abstract}

\maketitle

\section{Introduction}

The study of socioeconomic networks is a burgeoning field in the physics and
economics literature, with major progress having been attained over the last
decade \cite{Jackson96, Bala00, Marsili04, Cajueiro05, Ehrhardt06,
DeMartino06}. Individuals and firms interact through networks to share
information and resources, exchange goods and credit, make new friendships or
partnerships etc. The structure of the network through which interactions take
place may thus have an important effect on the success of the individual or
the productivity of the firm \cite{Jackson96}. Furthermore, the network of
interactions among socioeconomic agents plays an important role for the
stability and efficiency of socioeconomic systems \cite{Iori06}. Theories
about how interaction networks form are thus essential for a deeper
understanding of the development and organization of society as a whole.

The economics literature focuses mainly on equilibrium networks and the
network formation mechanisms are based on utility maximization and costs
minimization. The aim of most economic papers is to identify, among the set of
equilibrium networks, the geometry that optimizes efficiency \footnote{A
network $g$ is efficient with respect to an aggregate utility measure $u$ if
$u\left(  g\right)  \geq u\left(  g^{^{\prime}}\right)  $ $\forall
g^{^{\prime}}\in\mathcal{G}$ \cite{Jackson96}.} in the sense of social
benefit. Likewise, economists are interested in the stability \footnote{A
network is pairwise stable when no node would benefit from severing an
existing link, and no two nodes would benefit from forming a new link
\cite{Jackson96}. Pairwise stability is stronger than Nash equilibrium and is
aimed at sequential updating.} of equilibrium networks under link deletion,
addition or rewiring \cite{Jackson96, Bala00}. A shortcoming of these models
is that the equilibrium networks are often too simple in their geometry
(stars, complete networks, interlinked stars, etc.), typically as a
consequence of the symmetries that need to be assumed in the payoff functions
in order to make the models analytically tractable \cite{Jackson06}.

The physics literature, instead, has mainly focused on the characterization of
the structure of real networks and proposed dynamic models, mostly based on
probabilistic rules, capable of reproducing the observed geometrical
structures (Poisson, stretched exponential and scale-free networks)
\cite{Albert02, NewmanSIAM03, DorogovtsevBook03}.

In this paper we try to combine the physics and economic approaches, by
introducing a stochastic network formation mechanism inspired by economists'
utility maximization models, which naturally extends the well known
physicists' preferential attachment rule \cite{Barabasi99}.

One of the most interesting models of socioeconomic network formation was
introduced by Jackson and Wolinsky\ in $1996$ \cite{Jackson96}. In their
model, the formation and evolution of links is driven by a utility
maximization mechanism. The model is based on the assumption that agents may
derive benefit not only from the nodes to which they are directly connected
(their nearest neighbours), but also from the ones they are connected to
indirectly (possibly via long paths). Less distant connections are more
valuable than more distant ones, but connections to the nearest neighbours are
costly. The utility of node $i$ is defined as:%

\begin{equation}
u_{i}=w_{ii}+\sum_{j\neq i}w_{ij}\delta^{d_{ij}}-\sum_{j\in{\mathcal{V}}%
(i)}c_{ij} \label{Utility}%
\end{equation}
where the contribution to the utility of $i$ from $j$ may depend on the weight
$w_{ij}$ of the edge between $i$ and $j$ (or, alternatively, on the fitness of
node $j$); $0\leq\delta<1$ captures the idea that the utility gain from
indirect connections decreases with distance; $d_{ij}$ is the number of links
in the shortest path between $i$ and $j$ ($d_{ij}=\infty$ if there is no path
between $i$ and $j$); ${\mathcal{V}}(i)$ is the set of nearest neighbours of
$i$; and $c_{ij}$ are the (node specific) costs to establish a directed
connection between $i$ and $j$ \footnote{In \cite{Jackson96} costs are assumed
to be equally, or cooperatively, shared by $i$ and $j$, but extensions to the
non cooperative case have also been explored.}. Costs can also be
differentiated in costs of initially creating or maintaining an edge
\cite{Bala00}.

The papers by Jackson and Wolinsky \cite{Jackson96}, as well as the one by
Bala and Goyal \cite{Bala00}, are mainly concerned with stability and
efficiency of the network resulting from different dynamic updating rules. In
particular, Jackson and Wolinsky study pairwise stability when agents can only
update a link at a time (either delete it or create it), while Bala and Goyal
allow agents to rearrange all their connections at once. The updating is
deterministic in both models, and a new configuration is accepted only if it
increases the utility of the agent. These two papers show that the star
network is both efficient and stable for a wide range of the parameters when
$\delta=1$. Nonetheless, a multiplicity of network architectures exist in
\cite{Bala00} for $0<\delta<1$ which could be a strict Nash equilibria, and to
which the system may converge depending of the initial conditions. Feri
\cite{Feri07} has shown that for sufficiently large networks the star network
is stochastically stable for almost all the range of parameters, even for
$0<\delta<1$.

Here we focus on the connections model of Jackson and Wolinsky, i.e. the case
$w_{ij}=1$, $w_{ii}=0$ and $c_{ij}=c$. In this case, the utility can be
written as
\begin{equation}
u_{i}=\sum_{l=1}^{l_{max}^{(i)}}\sum_{\{k|d_{ik}=l\}}\delta^{l}-\sum
_{j\in{\mathcal{V}}(i)}c=\sum_{l=1}^{l_{max}^{(i)}}\delta^{l}z_{l}%
^{(i)}-cz_{1}^{(i)} \label{Utility1_JW}%
\end{equation}
where the sum in $l$ is over all shortest paths of length $l$ from node $i$,
the sum in $k$ is over all nodes whose shortest path from $i$ is $d_{ik}=l$,
$l_{max}^{(i)}$ is the path length of the node the furthest away from node
$i$, and $z_{l}^{(i)}$ is the number of $l$th-nearest neighbours of node $i$.
The utility of a node is expressed in (\ref{Utility1_JW}) as a weighted sum of
the number of nodes accessible from $i$ on outward "layers" of increasing
distance from $i$. Thus, we start at node $i$ and multiply $\delta$ by the
number of nodes that are joined by an edge to $i$--this being the first layer.
We then add $\delta^{2}$ times the number of nodes that are joined by an edge
to a node in the first layer--this is the second layer. We continue in this
way until no new nodes are found. Hence, expression (\ref{Utility1_JW})
incorporates implicitly the well known breath-first search algorithm
\cite{KleinbergBook06}. In this paper, we will consider only networks 
with zero costs. Therefore, equation (\ref{Utility1_JW}) becomes:
\begin{equation}
u_{i}=\sum_{l=1}^{l_{max}^{(i)}}\delta^{l}z_{l}%
^{(i)} \label{Utility1}%
\end{equation}

In the first part of the paper we focus on a specific network growth mechanism
and examine the resulting network topology. If each new node attached
deterministically to the existing node with maximal utility, the resulting
network would be a star. The randomness generated by a probabilistic attaching
rule can be interpreted as costs and barriers to gather information, or
bounded rationality, all of which limit the ability to establish links in an
optimal way, thus possibly generating more realistic geometries than the star
network. It is thus, worthwhile to ask which network topologies are to be
found when new nodes arrive steadily and create links with existing nodes in a
probabilistic way, proportionally to the utility of existing nodes. In this
way, we build on the preferential attachment growth rule of Barab\'{a}si and
Albert \cite{Barabasi99, Albert02} which can be recovered from equation
(\ref{Utility1}) when $l_{max}^{(i)}=1$. Furthermore, preferential attachment
is, arguably, the most extensively studied mechanism of network formation and
one that has revealed insights into properties observed in real networks.
Therefore, it is important to understand the robustness of the specific rule
of linear preferential attachment by node degree, which is one of the aims of
this paper.

Often, the specific network growth mechanisms are unknown
and only the topology of the equilibrium network can be extracted from data.
One obvious question is then how the observed equilibrium networks rank in
terms of their efficiency, e.g. Erd\H{o}s--R\'{e}nyi versus scale--free
networks. We address this question in the second part of the paper and derive analytical results by using the generating
function approach \cite{Wilf94}. We show that power-law networks are more
efficient than Poisson random network when individual utility is defined by
(\ref{Utility1}) thus providing a possible explanation for why scale-free
networks are so ubiquitous.

\section{Growing Networks\label{GrowthModel}}

\begin{figure}[ptb]
\begin{center}
\includegraphics[trim=0cm 0cm 0cm 0cm,scale=0.35]{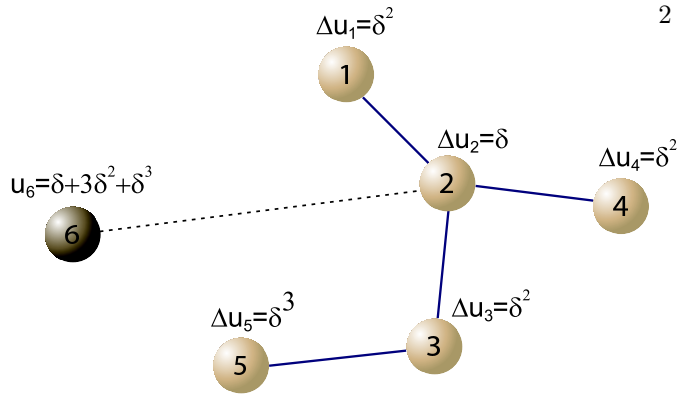}
\end{center}
\caption{Schematic layout of network growth when $m=m_{0}=1$. The addition of
a new node, $6$, implies an increase of the utility of nodes $1$ to $5$ which
is simply $\delta^{d}$, where $d$ is the path length from node $6$. The
simplicity of this updating mechanism allowed simulations to be run with
$N=10^{5}$ when $m=1$. }%
\label{Fig_network_dynamics}%
\end{figure}

\begin{figure*}[ptb]
\centering{
\includegraphics[trim=0cm 0cm 0cm 0cm,scale=0.7]{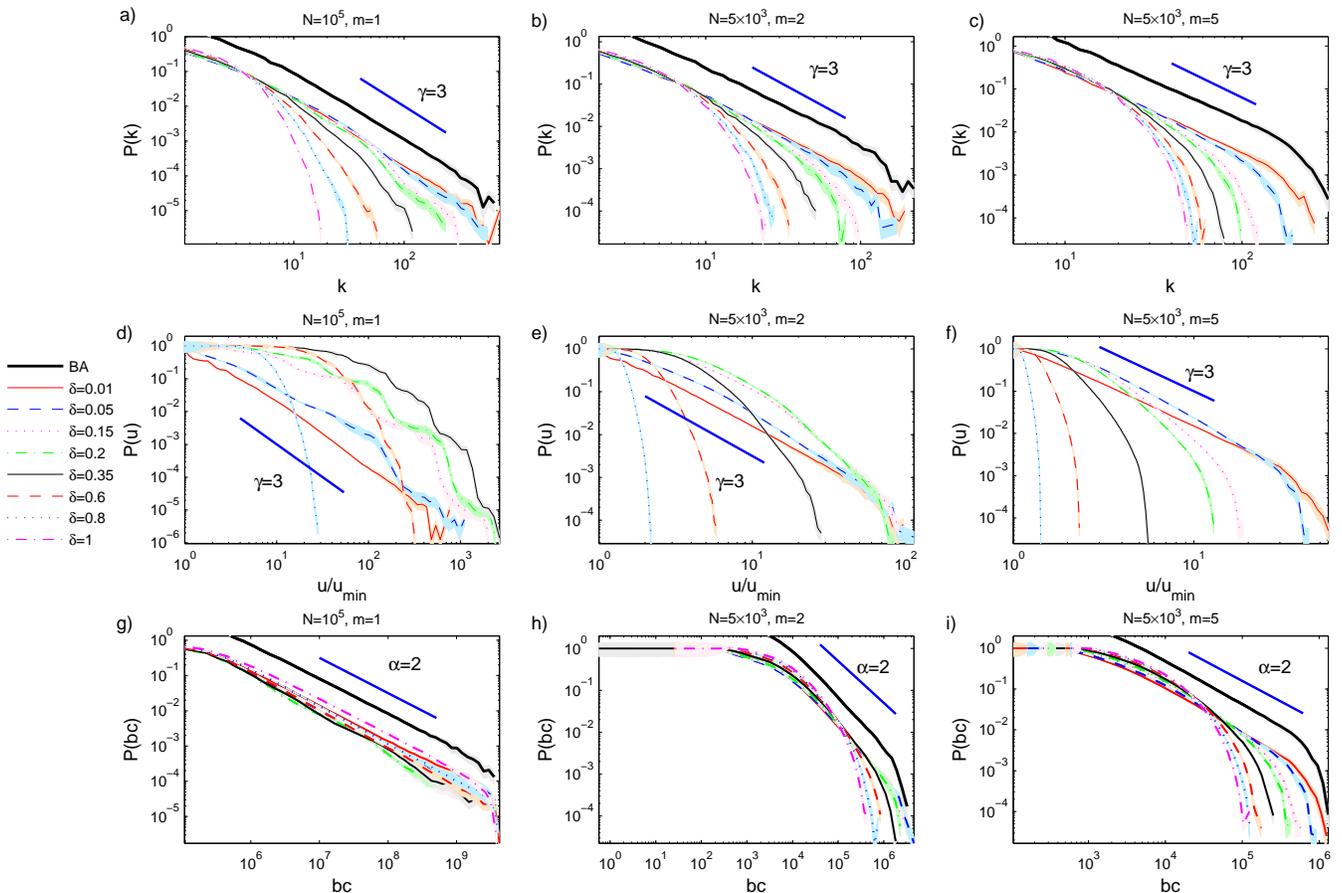}
}\caption{(Color online) Cumulative distribution function of degree (panels a), b) and c)),
utility (panels d), e) and f)) and betweenness centrality (panels g), h) and
i)) for several values of $\delta\in\left]  0,1\right]  $, and $m=1,2$ and
$5$. We also plot the corresponding distribution of degree and betweenness for
the BA model (curves were shifted vertically). Simulations were averaged over
$30$ runs in networks with $N=10^{5}$ ($m=1$) or $N=5\times10^{3}$ ($m=2$ and
$5$). Coloured bands around the curves are $95\%$ confidence intervals.}%
\label{Fig2}%
\end{figure*}

\begin{figure*}[ptb]
\centering
{\includegraphics[trim=0cm 0cm 0cm 0cm,scale=0.9]{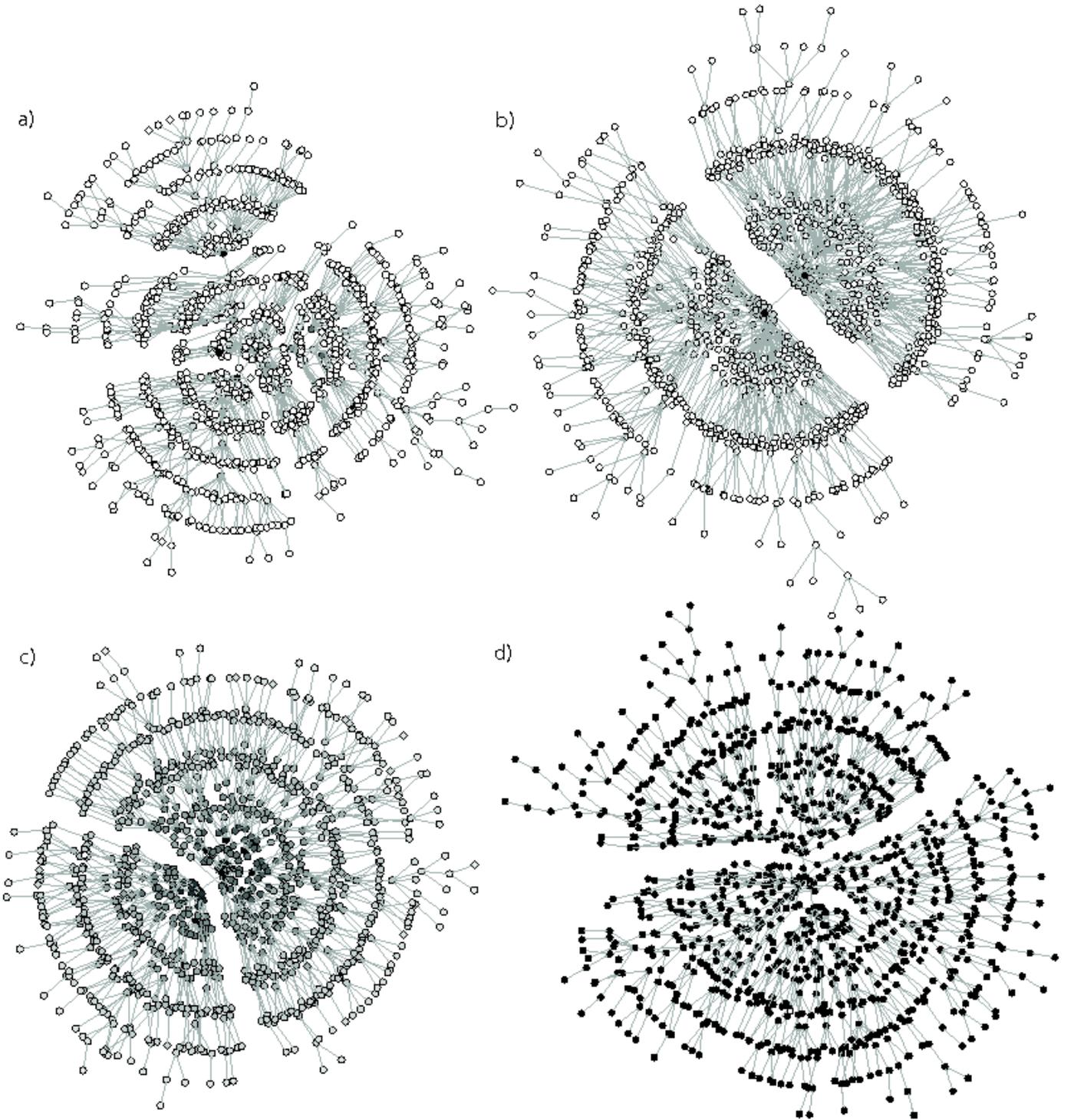} }%
\caption{Kamada-Kawai spring layout \cite{Kamada89} for $m=1$ and $N=10^{3}$.
Panel a) is a sample layout for $\delta=0.01$, b) for $\delta=0.2$, c) for
$\delta=0.7$ and d) for $\delta=1$. On each panel, nodes are coloured by their
utility on a gray scale from minimal (white) to maximal (black) utility.}%
\label{Fig_Layouts}%
\end{figure*}

\begin{figure*}[ptb]
\begin{center}
\includegraphics[trim=0cm 0cm 0cm 0cm,scale=0.7]{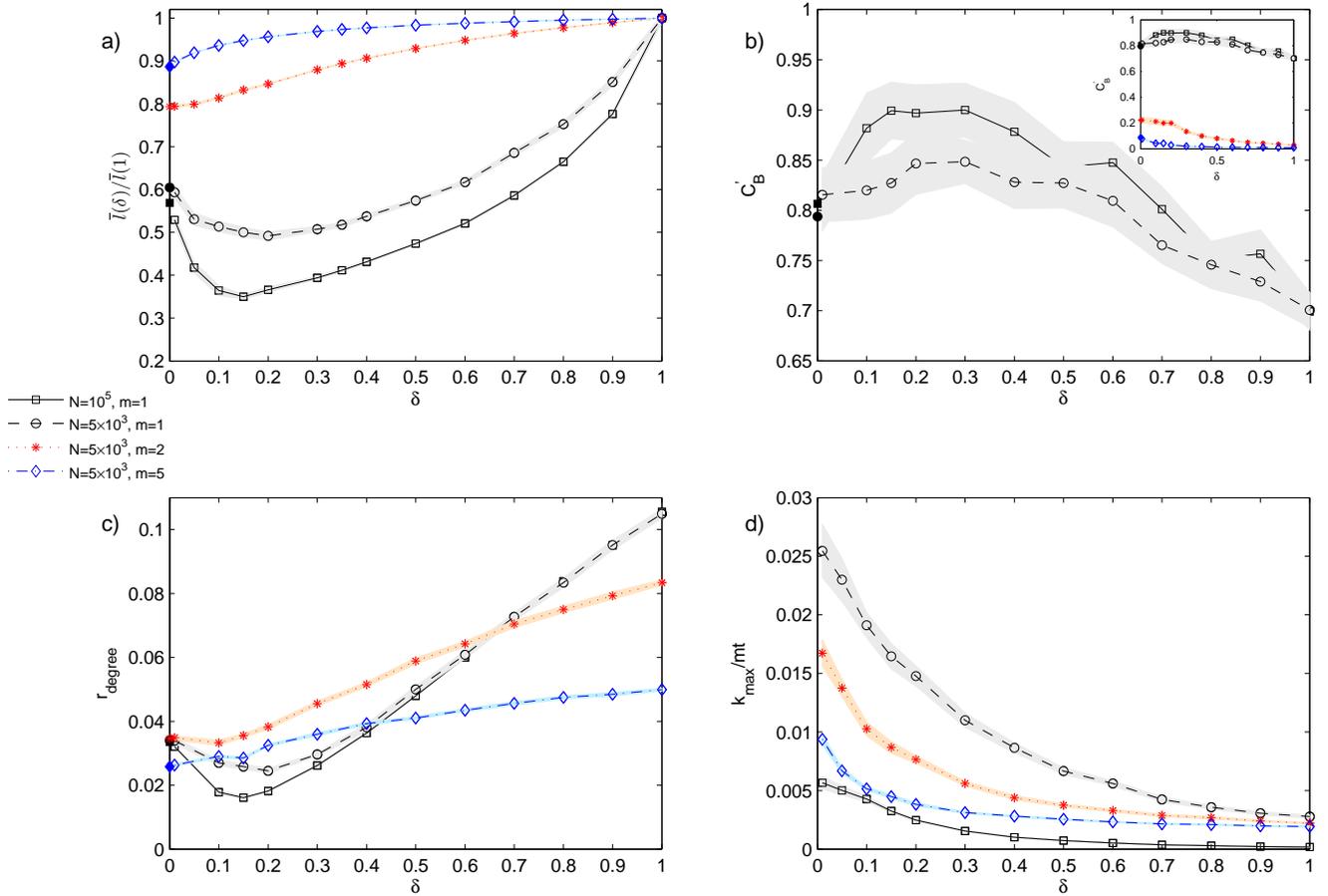}
\end{center}
\caption{(Color online) Plots of a) average path length, b) central point dominance, c) assortativity coefficient, and d) normalized maximum degree as a function of
$\delta$ for the simulation results when $m=1,2$ and $5$. Curves in panel a)
were scaled by the values at $\delta=1$ (exponential network). The BA values are indicated by the corresponding full symbols. Simulations were averaged over
$30$ runs and coloured bands around the curves are $95\%$ confidence intervals.}%
\label{Fig_average_path_length}%
\end{figure*}

\begin{figure}[ptb]
\begin{center}
\includegraphics[trim=0cm 0cm 0cm 0cm,scale=0.48]{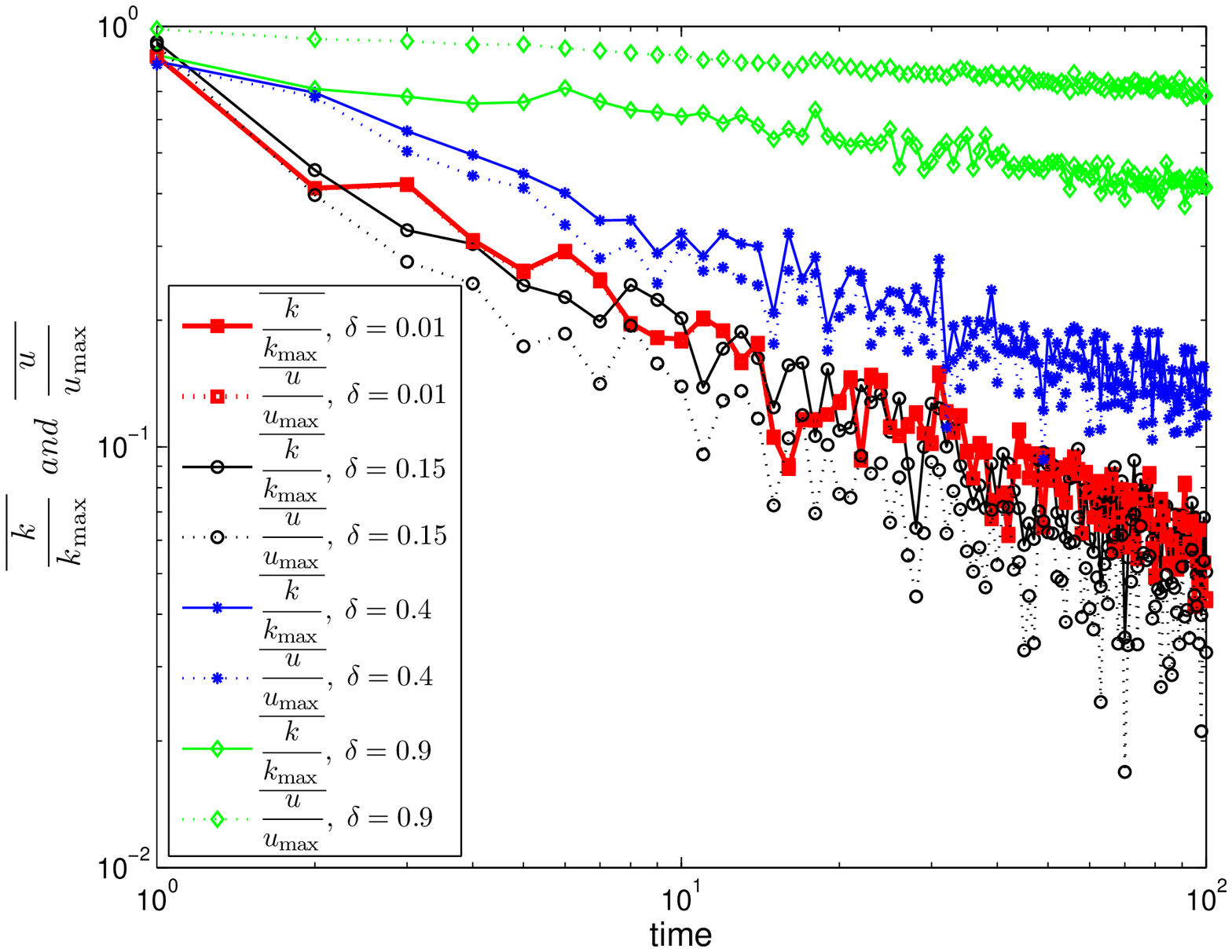}
\end{center}
\caption{(Color online) Plots of average $k/k_{max}$ and $u/u_{max}$ as a function of the
entry time for the first 100 nodes when $N=10^{5}$ and $m=1$. Simulations were averaged over $30$ runs.}%
\label{Fig_ratios}%
\end{figure}

In the classic Barab\'{a}si and Albert model \cite{Barabasi99}, a network is
grown by adding, at every time step, a new node that attaches to $m$ existing
nodes with a probability proportional to their degree, $\Pi(k_{i})=k_{i}
/\sum_{j=1}^{N}k_{j}$. At time $t$, the resulting network has size
$N_{t}=m_{0}+t$, where $m_{0}\geq m$ is the size of the (fully connected)
network at time $t=0$. 
Preferential attachment models were in fact introduced in the literature already by Yule \cite{Yule25} and Simon \cite{Simon55}.

The preferential attachment mechanism generates a scale-free
probability density of incoming links that leads to the stationary result
$p(k)=2m^{2}/k^{\gamma}$, with $\gamma=3$ independently of $m$.
The  model is also characterized by a clustering coefficient larger than the one found for the Erd\H{o}s R\'{e}nyi networks (for $m>1$) and no clear
assortative/disassortative behaviour \cite{Albert02}.

Several models have been proposed lately to investigate extensions of the
preferential attachment mechanism through edge removal and rewiring,
inheritance, redirection or copying; node competition, aging and capacity
constraints; and accelerated growth of networks to name just a few (see
\cite{Albert02, Dorogovtsev02, NewmanSIAM03, Durrett07} for reviews).

Of particular
relevance to our approach are fitness models \cite{Bedogne06, Caldarelli02,
Masi06, Servedio04}, where the probability of attaching to a node is
proportional to node fitness
\begin{equation}
\Pi(k_{i})\sim\frac{f_{i}k_{i}}{\sum_{j=1}^{N}f_{j}k_{j}}.
\label{fitness_model}%
\end{equation}

Here we extend the preferential attachment rule by introducing a growing
mechanism inspired on the work of Jackson and Wolinsky \cite{Jackson96}. Our
contribution is to propose preferential attachment by node utility. Thus, the
probability that a new node $j$ will be connected to an existing node $i$
depends on the utility of $i$, such that%
\begin{equation}
\Pi_{i}=\frac{u_{i}}{\sum_{k=1}^{N}u_{k}} \label{probability}%
\end{equation}
where the utility of node $i$, $u_{i}$, is given by (\ref{Utility1}). 
Attachment happens with uniform distribution for $\delta=1$ generating  an exponential
distribution of node degree \footnote{This is Model A in \cite{Barabasi99}.}.
While  the model is not defined for $\delta=0$, the preferential attachment rule (\ref{probability}) is invariant up to
multiplicative factors in (\ref{Utility1}), so for $\delta\neq0$ the
qualitative behaviour of the model remains unchanged if we define utility as%
\begin{equation}
u_{i}^{^{\prime}}=\frac{u_{i}}{\delta}=k_{i}+\sum_{l=2}^{l_{max}^{(i)}}%
\delta^{l-1} z_l^{(i)} \label{UtilityGeneralized}%
\end{equation}
where $k_{i}$ is the degree of node $i$. Thus, as $\delta\rightarrow0$ our
model converges to the Barab\'{a}si-Albert model and generates a   scale-free network. 

Our model has resemblances with the fitness models discussed above. However,
there is a fundamental discrepancy: we regard utility as a time-dependent
measure of node fitness, whereas existing models assume that node fitness does
not change with time.

At each time step, a new node $j$ joins the network and the utility of
existing nodes changes. When $m=1$, the utility increment to an existing node
$i$ at distance $l$ from $j$ is given by $\Delta u_{i}=\delta^{l}$ and
therefore, at each time step, the computation of utility for the network can
be completed in $O(N)$ time. Figure \ref{Fig_network_dynamics} is a diagram of
a possible network configuration with $m=m_{0}=1$ after $t=5$ time steps,
showing the change in utility of existing nodes, $\Delta u_{i}=\delta^{l}$.
When $m>1$, the increment in the utility of node $i$ depends on the existing
network geometry and $\Delta u_{i}\ge\delta^{l}$. Therefore, when $m>1$, we need
to re-compute the utility of \textit{all} existing nodes at every time step,
and the computation runs in $O(N^{2})$ time as it involves running a
\textit{breadth-first-search} algorithm from every node. This is the reason
why we have ran simulations for $N=10^{5}$ when $m=1$, but only up to
$N=5\times10^{3}$ when $m>1$.

Existing nodes $i$ at a higher distance than a certain $l_{\max}$ from new
node $j$ receive a contribution $\Delta u_{i}=\delta^{d(j,i)}<10^{-precison}$
which is less than the number of significant digits that the computer can
store (typically $precison=32$ in double precision), and do not need to have
their utility updated in the simulations. This maximal distance $l_{\max}$ is
defined as%

\begin{equation}
10^{-precision}>\delta^{l_{\max}}\Leftrightarrow l_{\max}>-\frac
{precision}{\log_{10}\delta}\label{precision}%
\end{equation}
Our implementation of the algorithm updates the utility of all nodes
accessible from the new node $j$ up to distance $l_{\max}=-32/\log_{10}\delta
$. The code was implemented in C++ and ran on a Condor framework (high
throughput computing) \cite{Thain05} for several values of $\delta$. Ensemble
averages were taken over $30$ runs.

Expressions (\ref{Utility1}) and (\ref{UtilityGeneralized}) predict the
existence of two distinct regimes: a scale-free regime as $\delta\rightarrow0$
($\delta\neq0$) and a random growth regime for which degree distribution is  exponential at $\delta=1$. We are
interested in exploring how the network evolves from one limit regime to the
other as we increase $\delta$.

Figure \ref{Fig2}a-c) shows the distribution of degree for $m=1,2$ and $5$.
We also plot the corresponding distribution for the BA model (solid curves
shifted vertically). For very small $\delta\sim0.01$, preferential attachment
by degree is indistinguishable from preferential attachment by utility and the
probability distribution of both quantities decay like $p\left(  x\right)
\sim x^{-\gamma}$ with $\gamma=3$. The power-law decay in the BA model is
known to be a peculiarity of the linear preferential attachment mechanism and
is destroyed by small perturbations like, for example, a non-linear attachment
rule $\Pi(k_{i})\sim k_{i}^{\alpha}$ \cite{Barabasi99}. Here we also observe a
depart from the scale-free regime as we increase $\delta$. Furthermore, in the
Barab\'{a}si-Albert model the degree distribution decays as a power-law with
exponent $\gamma=3$ independently of $m$. In our model, increasing $m$ has the
effect of homogenizing the utility of the nodes (the distance between pairs of
nodes decreases with increasing connectivity in the networks). Consequently,
deviations from the power-law decay are observed at lower and lower values of
$\delta$ as we increase $m$.

Betweenness centrality is plot in Figure \ref{Fig2}g-i) as $m$ is varied.
Recent results have shown that the distribution of loads (or betweenness)
scales with a power law \cite{Goh01, Barthelemy04} $p(g)\sim g^{-\alpha}$
where $\alpha=2$ for a tree (and hence for $m=1$). This justifies the collapse
of the curves of the distribution of betweenness in Figure \ref{Fig2}g).  As can be observed in Figure \ref{Fig2}h) and i), the
distribution of betweenness deviates from the power-law behaviour as $m$ is increased.

For intermediate values of $\delta$ and $m=1$ a number of interesting features
appear. First of all, we observe that the utility distribution becomes
step-like  (Figure \ref{Fig2}d) suggesting the presence of subsets of nodes
that share similar utilities.

This phenomenon can also be inferred from the network layouts in Figure
\ref{Fig_Layouts} (for networks of $10^{3}$ nodes with $m=1$), which are produced using the Kamada-Kawai spring layout \cite{Kamada89}. Essentially, the Kamada-Kawai layout assigns stronger springs to vertices that are closer in the graph-theoretic sense (i.e., by following edges) and therefore places them closer together. In the case $m=1$ (a tree), nodes close to the hubs on the graph layout will also be close 
in graph terms and therefore we can interpret the layout heuristically: these nodes have similar utility. When $\delta=0.01$ (close to the BA scale-free regime), the layout shows a few utility hubs (the dark
vertices) surrounded by clouds of nodes that disperse as we move further away
from the hubs; for $\delta=0.2$ denser clouds of nodes cluster around a
smaller number of hubs, and can still be observed farther away from the hub;
for higher $\delta$ the clouds start dispersing; eventually for $\delta=1$ all nodes have the same utility.

\begin{figure}[tbh]
\begin{center}
\includegraphics[width=8.558cm]{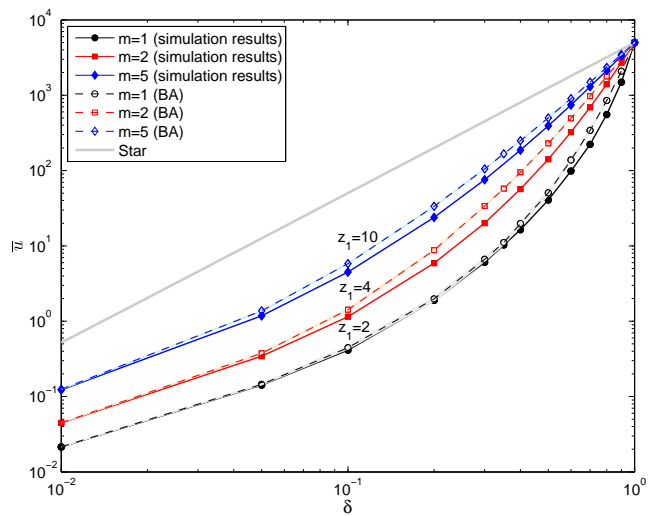}
\end{center}
\caption{(Color online) Average utility for the simulation results (solid curve and symbols)
and the BA model (dashed curve, open symbols) for $m=1$ ($z_{1}=2$), $2$
($z_{1}=4$) and $5$ ($z_{1}=10$) when $N=5\times10^{3}$. We also plot the average utility of a star with $N=5\times10^{3}$ nodes. Simulation results
were averaged over $30$ runs. Coloured bands around the curves are $95\%$
confidence intervals.}%
\label{Figure_utility_curves_simulation}%
\end{figure}

The rearrangement of the network as $\delta$ increases from zero gives rise to non monotonous behaviour of the average path length
$\bar{l}$, degree assortativity $r_{degree}$ \cite{NewmanPRE03}, and central
point dominance $C_{B}^{\prime}$ \cite{FreemanSociometry77}. Both $\bar{l}$ and $r_{degree}$ show a minimum for
$\delta\sim0.15$ ($N=10^{5}$ and $m=1$) and $C_{B}^{\prime}$
has a maximum around the same value. This behaviour can be observed in
Figure \ref{Fig_average_path_length}a-c). The average path length
$\overline{l}(\delta)$ is measured relative to the path length of a random
growing network (i.e. relative to $\overline{l}(1)$). The scale-free regime is characterised by a shorter path length than the
random growth regime. Here we observe an even further contraction of the network for
values of $\delta$ up to $\sim0.5$. Note that the average path length of a star
network is $l^{\ast}=2$ for $N$ large. When normalizing with
$l(1)=20.16$ we get a value of $l^{\ast}\sim0.1$. While this value is
still much smaller than $\overline{l}(0.15)$, the network contraction seems to
indicate a move toward a more star-like configuration. To further investigate
this point, we compute the the central point dominance measure introduced by
Freeman \cite{FreemanSociometry77} and plot it in Figure \ref{Fig_average_path_length}b). This measure is defined as the average
difference in node centrality (measured by node total betweenness) between the
most central point and all the others. The central point dominance takes a value
between zero (for a graph where all points have the same centrality) and one
(for the wheel or star graph). The maximum of $C_{B}^{\prime}$ for $\delta
\sim0.15$ in Figure \ref{Fig_average_path_length}b) confirms that the network
is becoming more star like around these values of $\delta$. 

Next, we plot
the assortativity of the network in Figure \ref{Fig_average_path_length}c). We
implement as measure of assortativity the degree assortativity $r_{degree}$  \cite{NewmanPRE03} which takes values from $-1$ to $+1$: negative values for disassortative networks,  $0$ if the networks are neither dissasortative or assortative, and $+1$
for fully assortative networks. This value approaches zero
for large $N$ in the BA model \cite{NewmanPRL02} and is negative for a star. Our model
shows a lower assortative mixing than the BA model for values of $\delta$ up to
$\sim0.5$. The decrease in $r_{degree}$ is also consistent with our hypothesis that the network is becoming more star like at $\delta\sim0.15$. While the  network goes through this rearrangement the degree of the most connected node (as a fraction of the total number of links) is nonetheless monotonously decreasing with $\delta$, as shown in Figure \ref{Fig_average_path_length}d).  The same behaviour  is observed for the utility of the most connected node (not shown here). This reveals that as new nodes are added to the network,  they do attach on average closer  to the hubs as $\delta$ increases, generating a more compact network,  but not directly  to them.

To gain further insights into the structural changes that take place as $\delta$ increases from zero, we analyze the role of entry time on node connectivity by
computing the ratios $u/u_{max}$ and $k/k_{max}$ for the first $100$ nodes as a function of the time at which they entered the network. The ratios plotted in
Figure \ref{Fig_ratios} are averaged over 30 different simulations, for 
networks with $N=10^{5}$ and $m=1$. The plot shows that for any $\delta$ the
initial node is likely to acquire the highest fraction of links and utility. Moreover, for
$\delta=0.15$ both the degree and utility ratios decay faster than in the
scale-free regime. This reveals that, for 
$\delta\sim0.15$, the earlier nodes receive both a higher relative degree and  utility than in the scale-free case. In other words, the earlier nodes are stronger hubs than in the scale-free case and thus the network arranges 
in a more star like configuration around $\delta=0.15$. As $\delta$ increases further, the slope of the utility ratio becomes lower than in the scale-free regime. In this
range, the utility differences between old and new nodes are not large enough
to create well defined utility or degree hubs. 

Figure \ref{Fig_ratios} also highlights the fundamental mechanism of structure formation in our model: the fine relation between degree and utility as $\delta$ is varied. In the scale-free regime, preferential attachment by utility is equivalent to preferential attachment by degree and node degree and utility assume the same value as the network grows. At $\delta\sim0.15$, we observe a gap between the two scaled quantities, suggesting a discrepancy between degree and utility for the higher order neighbours of the utility hubs. This gap is larger for $\delta=0.15$ than  $\delta=0.4$, indicating that 
around $\delta\sim0.15$ the growth mechanism is generated by a variable (node utility) which is considerably independent of node degree and thus revealing why this is the region where the network displays more interesting structure. As $\delta$ increases towards $1$, the influence of random network growth becomes more important and this structure generating mechanism disappears.

Finally, we investigate how average node utility compares in networks generated with our preferential utility attachment, the scale-free regime (here generated via the BA preferential attachment
mechanism) and a star network. The average utility of a star network is given by:
\begin{equation}
\overline{u}_{\ast}\left(  \delta\right)  =\delta z_{1}\left(  1+\delta
\frac{N-2}{2}\right)  \label{average_u_star}%
\end{equation}
where $z_{1}=2(N-1)/N$. For $N$ large, $z_{1}\simeq2$ and $\overline{u}_{\ast
}\left(  \delta\right)  \sim N\delta^{2}$.
In Figure \ref{Figure_utility_curves_simulation} we plot the average utility
for networks in our model at different $\delta$ when $N=5\times10^{3}$ and
$z_{1}=2$, $4$ and $10$ and compare that to the corresponding average utility in the BA model and a star with the same $N$. The plot shows that the BA scale-free network  has a higher utility than
the network generated via our preferential utility mechanism at all values of delta and $z_{1}$. Networks in our model become more star-like for $\delta\sim0.5$, but this implies an increase in the utility of only  a small number of nodes (the early nodes). Therefore, the average utility of nodes in the network is still higher in the BA model than in our model for the same values of $\delta$ due to the scale-free structure of the former.  Comparisons with the 
star network can  only be made  when $z_{1}=2$ as
this is the average degree of the star network when $N$ is large. Figure \ref{Figure_utility_curves_simulation}  confirms that the star network has the highest utility for this value of $z_{1}$ among all the networks we study.  Nevertheless, the star network can only be achieved if agents are perfectly rational and have access to full information (in which case the attachment mechanism would be deterministic). This is rarely the case in real word situation, thus the comparison with the star is of little practical relevance. 

\section{Analytical Results for Random Networks\label{AnalyticalResults}}

An interesting question to ask (for example, from the point of view of the
social planner) is how network topologies rank against each other and which
network structure maximizes the total, or the average utility (networks that
satisfy this condition are said to be efficient in economics). 
We show that it
is possible to derive analytical results for the average utility in Poisson and power-law networks. By comparing average utility in different
network topologies with the same size and the same average degree, we show
that power-law networks are more efficient than Poisson random networks (even
though less efficient than the star). The effect of costs on
$\overline{u}\left(  \delta\right) $ is a constant term for all networks with
the same $N$ and $z_{1}$. Therefore, without loss of generality, we choose
$c=0$ in the analysis below.

\begin{figure}[ptb]
\begin{center}
\includegraphics[height=5.3883cm,width=8.7008cm]{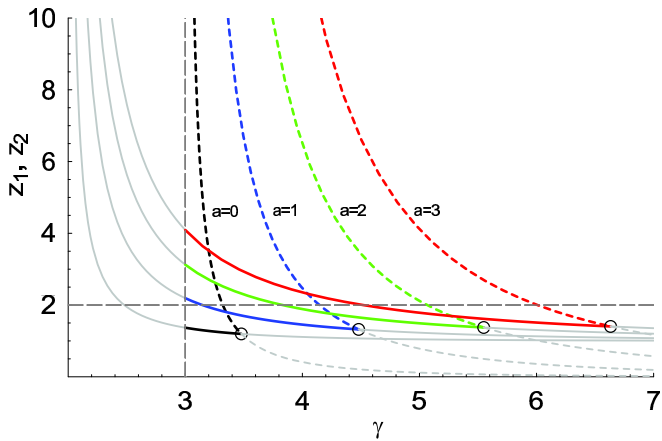}
\end{center}
\caption{(Color online) Average number of first and second-neighbours ($z_{1}\left(
\gamma,a\right)  $ and $z_{2}\left(  \gamma,a\right)  $) in networks with
degre distribution given by (\ref{prob_scale_free}). From left to right, we
plot $z_{1}\left(  \gamma,a\right)  $ (full curves) and $z_{2}\left(
\gamma,a\right)  $ (dashed curves) for $a=0$ (black), $1$ (blue), $2$ (green)
and $3 $ (red). The values of $z_{1}\left(  \gamma\leq3,a\right)  $ for which
$z_{2}\left(  \gamma,a\right)  $ is not defined are plot in grey, as well as
regions of the curves for which $z_{1}\left(  \gamma,a\right)  >z_{2}\left(
\gamma,a\right)  $. The circles denote the intersection of the two curves,
$z_{1}\left(  \gamma,a\right)  $ and $z_{2}\left(  \gamma,a\right)  $, for
each value of $a$.}%
\label{Figz1z2}%
\end{figure}

\begin{figure}[ptbh]
\begin{center}
\includegraphics[height=6.8535cm,width=8.558cm]{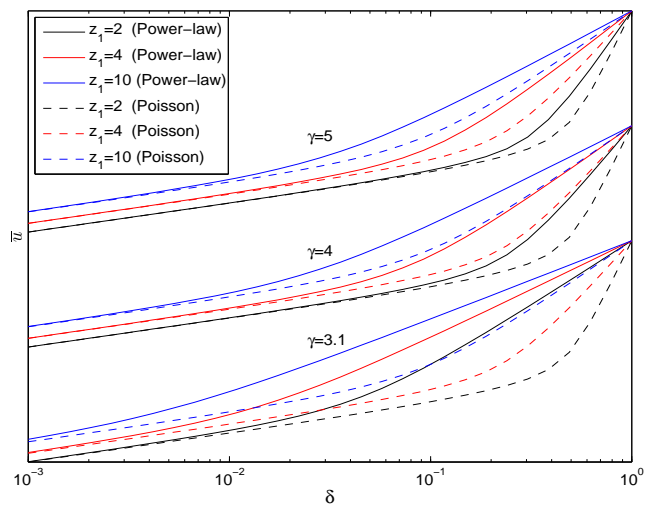}
\end{center}
\caption{(Color online) Analytical results for average utility in networks with power-law
(full curves) and Poisson (dashed curves) degree distributions as a function
of $\delta$, $z_{1}$ and $\gamma$ for $N=10^{5}$. Curves have been shifted
vertically for different values of $\gamma$ for clarity. Values of $z_{1}$
increase from bottom to top.}%
\label{Figure_utility_curves}%
\end{figure}

If the sum in (\ref{Utility1}) was to be evaluated up to distance
$l_{max}^{(i)}=1$ for every node, expression (\ref{average_U}) would simplify
to $\overline{u}\left(  \delta\right)  =\delta z_{1}$, \textit{i.e.} average
utility would be independent of the specific network topology and all networks
with the same number of nodes and links would be equally efficient. Thus we
need to introduce long range interactions ($l_{max}^{(i)} >1$) to be able to
rank networks in terms of their efficiency. 

To derive an expression for average utility in generic random networks with $N$ large, we
average both sides of (\ref{Utility1}):
\begin{equation}
\overline{u}\left(  \delta\right)  =\sum_{l=1}^{\overline{l}}\delta^{l}
z_{l} \label{average_U}
\end{equation}
where $z_{l}$ is the average number of $l$th neighbours of a node.  Newman \textit{et al.} \cite{Newman01, Newman02} define $\overline{l}$  via the expression 
\begin{equation}
1+\sum_{l=1}^{\overline{l}}z_{l}=N \label{average path length condition}%
\end{equation}

Now that we have expressed average utility in terms of the breadth-first
search algorithm, we can derive a closed form of expression (\ref{average_U})
if we have access to analytical expressions for $\overline{l}$ and $z_l$. This can be accomplished by generating
functions, which are particularly useful when determining means, standard
deviations and moments of distributions \cite{Wilf94}.

The average number of neighbours (average degree) and the average number of
second neighbours of a node can be derived from the probability generating
function of node degree, $G_{0}\left(  x\right)  =\sum_{k=0}^{\infty}%
p_{k}x^{k}$, as long as the degree distribution, $p_{k}$, is specified. The
beauty of the generating function formalism is that one can derive $z_{l}$ as
a function of $z_{1}$ and $z_{2}$ only \cite{Callaway00, Newman01, Newman02}:%

\begin{equation}
z_{l}=\left[  \frac{z_{2}}{z_{1}}\right]  ^{l-1}z_{1} \label{Zm}%
\end{equation}
Replacing equation (\ref{Zm}) in equation (\ref{average_U}) yields%

\begin{equation}
\overline{u}\left(  \delta\right)  =\delta z_{1}\sum_{l=1}^{\overline{l}%
}\left(  \delta Z\right)  ^{l-1}=\frac{\delta z_{1}(\left(  \delta Z\right)
^{\overline{l}}-1)}{\delta Z-1} \label{averageutility}%
\end{equation}
where $Z=z_{2}/z_{1}$. For $Z>1
$ and $N>z_{1}+1$, which are conditions satisfied by most networks,
$\overline{l}$ can be calculated as a function of $N$, $z_{1}$ and $z_{2}$
from (\ref{average path length condition}) and (\ref{Zm}) as \cite{Newman01}:%
\begin{equation}
\overline{l}=\frac{\ln[\left(  N-1\right)  \left(  Z-1\right)  /z_{1}+1]}%
{\ln\left(  Z\right)  } \label{avrg path length}%
\end{equation}
In what follows, we investigate the behaviour of (\ref{averageutility}) for
Poisson and power-law random networks.

Poisson random networks are characterized by $z_{1}=pN$ and $z_{2}=z_{1}^{2}$
\cite{Newman01}, thus (\ref{avrg path length}) yields $\overline{l_{P}}%
=\ln(\frac{N(z_{1}-1)+1}{z_{1}})/\ln(z_{1})$. In this case,
(\ref{averageutility}) becomes:%
\begin{equation}
\overline{u}_{P}\left(  N,\delta,z_{1}\right)  =\frac{z_{1}\delta\left(
\left(  \delta z_{1}\right)  ^{\ln(N+\frac{1-N}{z_{1}})/\ln(z_{1})}-1\right)
}{\delta z_{1}-1} \label{U_poisson}%
\end{equation}
for $N>z_{1}+1$, $0<\delta\leq1$ and $z_{1}>1$.

Next, we consider power-law networks with degree distribution of the form:%
\begin{equation}
p_{k}\left(  \gamma,a\right)  =\frac{1}{\zeta\left(  \gamma,1+a\right)
}(a+k)^{-\gamma}\quad,a\geq0 \label{prob_scale_free}%
\end{equation}
where the normalizing factor $\zeta\left(  \gamma,a+1\right)  =\sum
_{k=1}^{\infty}(a+k)^{-\gamma}$ is the Hurwitz zeta function ($\gamma>1$). The
generating function for the probability distribution is given by%
\begin{equation}
G_{0}\left(  x,\gamma,a\right)  =\sum_{k=1}^{\infty}p_{k}x^{k}=\frac
{x\Phi(x,\gamma,a+1)}{\zeta\left(  \gamma,a+1\right)  } \label{G0_scale_free}%
\end{equation}
where $\Phi(x,\gamma,a)=\sum_{k=0}^{\infty}\frac{x^{k}}{(a+k)^{\gamma}}$ is
the Lerch transcendent. For our purposes, only the first two derivatives of
$\Phi(x,\gamma,a+1)$ with respect to $x$ are relevant as the average number of
first and second-neighbours are given, respectively, by $z_{1}\left(
\gamma,a\right)  =\left.  \frac{\partial G_{0}(x)}{\partial x}\right\vert
_{x=1}$ and $z_{2}\left(  \gamma,a\right)  =\left.  \frac{\partial^{2}%
G_{0}(x)}{\partial x^{2}}\right\vert _{x=1}$. Hence \begin{widetext}
\begin{align}
z_{1}\left(  \gamma,a\right)    & =\frac{\Phi(1,\gamma-1,a+1)-a\Phi
(1,\gamma,a+1)}{\zeta(\gamma,a+1)}\quad,\gamma>2\wedge a\geq
0\label{z1_scale_free}\\
z_{2}\left(  \gamma,a\right)    & =\frac{\zeta(\gamma-1,a+1)}{\zeta
(\gamma,a+1)}z_{1}\left(  \gamma-1,a\right)  -(a+1)z_{1}\left(  \gamma
,a\right)  \quad,\gamma>3\wedge a\geq0\label{z2_scale_free}%
\end{align}
Thus
\begin{equation}
Z\left(  \gamma,a\right)  =\frac{\zeta(\gamma-1,a+1)}{\zeta(\gamma,a+1)}%
\frac{z_{1}\left(  \gamma-1,a\right)  }{z_{1}\left(  \gamma,a\right)
}-a-1\quad,\gamma>3\wedge a\geq0\label{Z_scale_free}%
\end{equation}
Substituting (\ref{z1_scale_free}) and (\ref{Z_scale_free}) into
(\ref{avrg path length}), we find%
\begin{equation}
\overline{l}_{SF}\left(  N,\gamma,a\right)  =\frac{\ln\left(  -\frac{(a+2)(N-1)}%
{z_{1}(\gamma,a)}+\frac{z_{1}(\gamma-1,a)\zeta(\gamma-1,a+1)(N-1)}%
{z_{1}(\gamma,a)^{2}\zeta(\gamma,a+1)}+1\right)  }{\ln\left(  -a+\frac
{z_{1}(\gamma-1,a)\zeta(\gamma-1,a+1)}{z_{1}(\gamma,a)\zeta(\gamma
,a+1)}-1\right)  }\quad,N>z_{1}\left(  \gamma,a\right)  +1\wedge
\gamma>3\wedge a\geq0\wedge Z\left(  \gamma,a\right)
>1\label{avrg path length scalefree}%
\end{equation}
and thus average utility is given by
\begin{equation}
\overline{u}_{SF}\left(  N,\delta,\gamma,a\right)  =\frac{\delta z_{1}\left(
\gamma,a\right)  \left(  \left(  \delta Z\left(  \gamma,a\right)  \right)
^{\overline{l}_{SF}\left(  N,\gamma,a\right)  }-1\right)  }{\delta Z\left(
\gamma,a\right)  -1}\quad,N>z_{1}\left(  \gamma,a\right)  +1\wedge
0<\delta\leq1\wedge
\gamma>3\wedge a\geq0\wedge Z\left(  \gamma,a\right)
>1\label{U_avrg_scale_free}%
\end{equation}
\end{widetext}where $z_{1}\left(  \gamma,a\right)  $, $Z\left(  \gamma
,a\right)  $ and $\overline{l_{SF}}\left(  N,\gamma,a\right)  $ are given by
(\ref{z1_scale_free}), (\ref{Z_scale_free}) and
(\ref{avrg path length scalefree}), respectively.

When $a=0$, the distribution of degree, (\ref{prob_scale_free}), becomes a
pure power-law $p_{k}\left(  \gamma\right)  =\frac{1}{\zeta\left(
\gamma\right)  }k^{-\gamma}$. In this case, we have $\left.  \zeta\left(
\gamma,a+1\right)  \right\vert _{a=0}=\zeta\left(  \gamma\right)  $ and
$\left.  \Phi(x,\gamma,a+1)\right\vert _{a=0}=\frac{Li_{\gamma}\left(
x\right)  }{x}$, therefore (\ref{G0_scale_free}) becomes%

\begin{equation}
G_{0}\left(  x,\gamma\right)  =\frac{\text{Li}_{\gamma}\left(  x\right)
}{\zeta\left(  \gamma\right)  } \label{G0_scale_free_a=0}%
\end{equation}
This generating function is also obtained for the power-law distribution with
exponential cut-off, proposed in \cite{Aldana03, NewmanPRL05}, $p_{k}\left(
\gamma,\kappa\right)  =Ck^{-\gamma}e^{-k/\kappa}$, in the limit $\kappa
\rightarrow\infty$.

Expression (\ref{G0_scale_free_a=0}) implies%

\begin{equation}
\left.  z_{1}\left(  \gamma\right)  \right\vert _{a=0}=\frac{\zeta\left(
\gamma-1\right)  }{\zeta\left(  \gamma\right)  }\quad,\gamma>2
\label{z1_scale_free_a=0}%
\end{equation}%
\begin{equation}
\left.  z_{2}\left(  \gamma\right)  \right\vert _{a=0}=\frac{\zeta\left(
\gamma-2\right)  -\zeta\left(  \gamma-1\right)  }{\zeta\left(  \gamma\right)
}\quad,\gamma>3 \label{z2_scale_free_a=0}%
\end{equation}
Therefore, in pure power-law networks, when $N\rightarrow\infty$, the average
number of second-neighbours, $z_{2}\left(  \gamma\right)  $, is finite only
for $\gamma>3$. However, the Riemann zeta function, $\zeta\left(
\gamma\right)  $, is a decreasing function of $\gamma$ (for $\gamma>3$) and
$z_{1}\left(  \gamma=3\right)  =\pi^{2}/6\zeta(3)\simeq1.36843$. In other
words, the existence of $z_{2}\left(  \gamma\right)  $ implies $z_{1}\left(
\gamma\right)  <z_{1}\left(  \gamma=3\right)  \simeq1.36843$, which is a
non-realistically low value for average degree in real networks. This explains
why we have chosen the modified power-law distribution (\ref{prob_scale_free}).

The generating function (\ref{G0_scale_free}) encapsulates \textit{all} the
moments of the degree distribution \cite{Newman01}. Hence, the expressions for
$z_{1}\left(  \gamma,a\right)  $ and $z_{2}\left(  \gamma,a\right)  $,
(\ref{z1_scale_free}) and (\ref{z2_scale_free}), are only exact in the limit
$N\rightarrow\infty$. Further, $\overline{l}_{SF}\left(  N,\gamma,a\right)  $
and $\overline{u}_{SF}\left(  N,\delta,\gamma,a\right)  $, both of which
depend on $z_{2}\left(  \gamma,a\right)  $, are only defined where
$z_{2}\left(  \gamma,a\right)  $ is finite, \textit{i.e. }for $\gamma>3$.
Therefore, it is essential to understand the behaviour of $z_{1}\left(
\gamma,a\right)  $ and $z_{2}\left(  \gamma,a\right)  $ in power-law networks.
Figure \ref{Figz1z2} shows $z_{1}$ (full curves) and $z_{2}$ (dashed curves)
within the range $\gamma>3\wedge Z>1$ (where $\overline{l}_{SF}\left(
N,\gamma,a\right)  $ is defined) for, from left to right, $a=0,1,2$ and $3$.

Having deduced closed-form expressions for average utility in Poisson and
power-law networks, we can now compare both networks under the condition that
$z_{1}$ is the same. Figure \ref{Figure_utility_curves} is a plot of average
utility versus $\delta$ when $z_{1}=\left\{  2,4,10\right\}  $ and $N=10^{5}$
for Poisson and power-law networks. The average utility of Poisson networks is
completely specified by $N,\delta$ and $z_{1}$, but power-law networks defined
by (\ref{prob_scale_free}) have one extra degree of freedom in $z_{1}\left(
\gamma,a\right)  $. In this case, we compute $z_{1}$ numerically by solving
(\ref{z1_scale_free}) for $z_{1}\left(  \gamma,a\right)  =\left\{
2,4,10\right\}  $ when $\gamma=\left\{  3.1,4,5\right\}  $.
For all cases studied power-law networks are more efficient than Poisson networks.

\section{Discussion}

The growth mechanism we have proposed is a natural extension of the
Barab\'{a}si-Albert preferential attachment by degree to preferential
attachment by node utility. Our analysis shows that for small values of
$\delta$, the utility decay parameter, the network retains a scale-free
structure that is nonetheless destroyed when $\delta$ increases. We have
identified a regime in $\delta$ where the network is
characterized by a lower average path length and assortativity coefficient and
a higher central point dominance than the scale-free network. In this regime,
the distribution of utility is a step-like function and the network has a more star-like structure.

The derivation of analytical expressions for average utility in Poisson and
power-law networks reveals that the latter have higher $\overline{u}$ for the
range of parameters that is of significance in real-world networks ($z_{1}%
\geq2$). This suggests a novel mechanism which may explain the ubiquitous
presence of power-law networks, in particular in situations where
collaboration, interaction and information sharing among the nodes are of
paramount relevance.
\\
\\

Social networks are highly volatile. Friendships can be  stable for a long time but  occasional encounters may lead to creation of links that are  never  used again in the future. A  dynamical model of network formation, where links not only can be heterogenous,  but whose  weights  can  change continuously over time,  would be a  more appropriate way to describe 
social interactions. Our assumption that $w_{ij}=1$ for all links is obviously a first order approximation and preferential growth, with no link rearrangements, 
is a crude description of social network formation. Nonetheless even  this simple mechanism can highlight  surprising features of the models (like in our case  a smaller  network diameter for intermediate values of $\delta$) and as such it is worth to investigate  in more general contests than the original BA model.

We have also assumed that the
connection costs in our model are zero. This  assumption is justified by the
fact that if costs are node independent they do not play any role in the
growing model. Similarly, costs do not play a significative role if we
restrict the comparison of average utility in section \ref{AnalyticalResults}
to networks with the same size and the same average degree. Nonetheless in a more realistic model, where links can be rearranged over time, costs would also play an important role in determining the shape of the network. 

Further analysis  taking into account both these effects is currently under development.

\begin{acknowledgments}
We wish to thank Ginestra Bianconi, Anirban Chakraborti, Francesco Feri,
Sanjeev Goyal, Matthew Jackson, Jukka-Pekka Onnela, and Marco Patriarca 
for stimulating discussions. We are grateful to the organizers and
participants of the EXYSTENCE Thematic Institute "From Many-Particle Physics
to Multi-Agent Systems" held at the Max Planck Institute for the Physics of
Complex Systems (MPIPKS) in Dresden, where the collaboration leading to this
work was started. RC would like to acknowledge support from the EPSRC
Spatially Embedded Complex Systems Engineering Consortium grant EP/C513703/1.
\end{acknowledgments}

\bigskip

\bibliographystyle{aip}
\bibliography{Giulia}

\end{document}